# Superconductivity in CuAl$_2$-type Co$_{0.2}$Ni$_{0.1}$Cu$_{0.1}$Rh$_{0.3}$Ir$_{0.3}$Zr$_2$ with a high-entropy-alloy transition metal site

Yoshikazu Mizuguchi[a]*, Md. Riad Kasem[a], and Tatsuma D. Matsuda[a]

[a]Department of Physics, Tokyo Metropolitan University, 1-1, Minami-osawa, Hachioji, 192-0397

**Abstract**

Research on high-entropy-alloy (HEA) superconductors is a growing field in material science. In this study, we explored new HEA-type superconductors and discovered a CuAl$_2$-type superconductor Co$_{0.2}$Ni$_{0.1}$Cu$_{0.1}$Rh$_{0.3}$Ir$_{0.3}$Zr$_2$ with a HEA-type transition metal site. A superconducting transition was observed at 8.0 K after electrical resistivity, magnetization, and specific heat measurements. The bulk characteristics of the superconductivity were confirmed through the specific heat measurements. The discovery of superconductivity in HEA-type Co$_{0.2}$Ni$_{0.1}$Cu$_{0.1}$Rh$_{0.3}$Ir$_{0.3}$Zr$_2$ will provide a novel pathway to explore new HEA-type superconductors and investigate the relationship between the mixing entropy and superconductivity of HEA-type compounds.

**Impact statement:** We report on the material design, synthesis, and observation of a superconducting transition at 8.0 K in new high-entropy-alloy-type compound with a CuAl$_2$-type structure, Co$_{0.2}$Ni$_{0.1}$Cu$_{0.1}$Rh$_{0.3}$Ir$_{0.3}$Zr$_2$.

# 1. Introduction

Recently, high-entropy alloys (HEAs) [1,2], which are defined as alloys containing five or more elements with a concentration between 5 to 35at%, have been extensively studied in the fields of material science, engineering, chemistry, and physics. Alloys synthesized to fit the criterion result in high configurational mixing entropy ($\Delta S_{mix}$), which is defined as $\Delta S_{mix} = -R \Sigma_i c_i \ln c_i$, where $c_i$ and $R$ are the compositional ratio and the gas constant, respectively [2]. In a HEA, five or more mixed elements are sometimes distributed randomly, and mixing elements sometimes results in the formation of nanoscale phase separations [3,4]. Such unique structural characteristics of HEAs have fascinated researchers in the field of superconductivity because nanoscale structural and/or electronic disorders could be useful for improving critical current density in some cases [5–7]. Recently, the exploration and investigation of HEA superconductors has been a hot topic since the discovery of the first HEA superconductor $Ta_{34}Nb_{33}Hf_8Zr_{15}Ti_{11}$ in 2014 with a transition temperature $T_c$ = 7.3 K [8,9]. After the discovery, various HEA superconductors with a simple alloy-type structure (bcc and hcp structures) have been discovered [9–15]. Although the pairing mechanisms of superconductivity in these HEA superconductors have been characterized as a conventional type, the field has been getting attention because HEA superconductors possess exotic characteristics. For example, electrical resistance measurements under extremely high pressures revealed that the superconductivity states in Ta-Nb-Hf-Zr-Ti are robust under pressures up to 190 GPa [16]. This fact suggests that the HEA states may be useful for maintaining the essential crystal structure and electronic states for the emergence of superconductivity under extreme conditions. In addition, as mentioned above, the concept of HEAs should be useful for research on improving the critical current density ($J_c$) of superconductors for practical use. Therefore, we need to extend the concept of HEAs to various compounds other than bcc and hcp metals to develop the field of HEA-type superconductors.

In 2018, Stolze et al. reported superconductivity in CsCl-type $(Sc,Zr,Nb,Ta)_{0.65}(Rh,Pd)_{0.35}$ [17]. Because a CsCl-type structure is composed of two different crystallographic sites, the $(Sc,Zr,Nb,Ta)_{0.65}(Rh,Pd)_{0.35}$ superconductor can be regarded as a *HEA-type compound*. Recently, we reported the synthesis and superconducting properties of HEA-type layered superconductors [18,19]. In a BiS$_2$-based layered superconductor system ($REO_{0.5}F_{0.5}BiS_2$: $RE$ = La, Ce, Pr, Nd, Sm), an increase in $\Delta S_{mix}$ improved the superconducting properties, owing to the suppression of local structural disorder [20,21]. Single crystal growth and the superconducting properties of HEA-type $RE(O,F)BiS_2$ have also been reported; superconducting properties comparable to low-entropy systems were confirmed [22]. In a $RE$123-type cuprate system with $RE$ = Y, La, Pr, Nd, Sm, Eu, Gd, no degradation of $T_c$ after an increase in $\Delta S_{mix}$ was confirmed [19]. Therefore, the HEA effects in layered superconductors seem to be working positively or at least less effectively. Recently, superconductivity in NaCl-type tellurides and related chalcogenides have been reported [23–25]. For example, a HEA-type telluride AgInSnPbBiTe$_5$ contains an HEA-type $M$ site ($M$ = Ag, In, Sn, Pb, Bi) and a Te site and exhibits superconductivity with $T_c$ = 2.6 K. In contrast to the layered systems, the superconducting properties of NaCl-type HEA tellurides are lower than those of low-entropy tellurides [24]. These results imply that the crystal structure type and its dimensionality are related to the effects of high-entropy alloying on the superconducting properties of compounds.

We explored a new HEA-type compound in this study. We have focused on the CuAl$_2$-type ($I4/mcm$) structure because over 100 superconductors could be found in a superconductor database (SuperCon, NIMS database) [26]. Among them, RhZr$_2$ has the highest $T_c$ of 11.3 K [27]. Furthermore, as summarized in Table I, all of the $Tr$Zr$_2$ ($Tr$ = Co, Ni, Rh, Ir) compounds exhibit superocnductivity with $T_c$ = 5.5, 1.6, 11.3, and 7.5 K, respectively [27]. Although FeZr$_2$ is a superconductor, its $T_c$ is lower than 1 K [28,29]. In the case of $Tr$ = Cu, CuZr$_2$ has a CuZr$_2$-type ($I4/mmm$) structure, but partial substitution of Cu for the CuAl$_2$-type compounds is possible and positively affects $T_c$. In Co$_{1-x}$Cu$_x$Zr$_2$, the $T_c$s for $x$ = 0.05 and 0.1 are higher than those for $x$ = 0 [30]. According to the $T_c$ and the crystal structure of the $Tr$Zr$_2$ compounds, we have designed a new HEA-type compound Co$_{0.2}$Ni$_{0.1}$Cu$_{0.1}$Rh$_{0.3}$Ir$_{0.3}$Zr$_2$, in which the composition at the $Tr$ site meets the compositional criterion of the HEA and achieves $\Delta S_{mix}$ ~ 1.5$R$ for the $Tr$ site.

**Table 1. Crystal structure type, space group, and $T_c$ of $Tr$Zr$_2$ and related phases ($Tr$ = Co, Ni, Cu, Rh, Ir).**

| Phase | Structural type | Space group | $T_c$ (K) | Reference ($T_c$) |
|---|---|---|---|---|
| CoZr$_2$ | CuAl$_2$-type | $I4/mcm$ (No. 140) | 5.5–6.0 | [27, this work] |
| NiZr$_2$ | CuAl$_2$-type | $I4/mcm$ (No. 140) | 1.6 | [27] |
| RhZr$_2$ | CuAl$_2$-type | $I4/mcm$ (No. 140) | 11.3 | [27] |
| IrZr$_2$ | CuAl$_2$-type | $I4/mcm$ (No. 140) | 7.5 | [27] |
| FeZr$_2$ | CuAl$_2$-type | $I4/mcm$ (No. 140) | < 1 K | [28,29] |
| CuZr$_2$ | CuZr$_2$-type | $I4/mmm$ (No. 139) | - | - |
| Co$_{0.9}$Cu$_{0.1}$Zr$_2$ | CuAl$_2$-type | $I4/mcm$ (No. 140) | 6.1 | [30] |
| Co$_{0.2}$Ni$_{0.1}$Cu$_{0.1}$Rh$_{0.3}$Ir$_{0.3}$Zr$_2$ | CuAl$_2$-type | $I4/mcm$ (No. 140) | 7.8 | This work |

## 2. Methods

A polycrystalline sample of Co$_{0.2}$Ni$_{0.1}$Cu$_{0.1}$Rh$_{0.3}$Ir$_{0.3}$Zr$_2$ was synthesized through arc melting in an Ar atmosphere. Powders of pure metals, Co (99%), Ni (99.9%), Cu (99.9%), Rh (99.9%), and Ir (99.9%), were mixed with a certain composition and pelletized. The metal pellet and a plate of pure Zr (99.2%) were used as starting materials for arc melting. The arc-melting was repeated after turning over the sample three times to homogenize the sample. The obtained sample was characterized using energy dispersive X-ray fluorescence analysis on a JSX-1000S (JEOL). The phase purity and crystal structure were examined through X-ray diffraction (XRD) with Cu-Kα radiation on a Miniflex-600 (RIGAKU) equipped with a high-resolution semiconductor detector D/tex-Ultra. The obtained XRD pattern was refined by the Rietveld method using RIETAN-FP [31], and schematic images of the refined crystal structure were depicted using VESTA [32]. Magnetization was measured using a superconducting quantum interference device (SQUID) on an MPMS-3 (Quantum Design). The temperature dependence of magnetization was measured after zero-field cooling (ZFC) and field cooling (FC). The susceptibility data shown in Fig. 2 was corrected using a demagnetization factor. Meanwhile, the magnetic field dependence of magnetization was measured from -7 to 7 T. The temperature dependence of electrical resistivity under magnetic fields up to 3 T was measured using a four-terminal method with a DC current of 5 mA on a GM refrigerator system (AXIS). The temperature dependence of specific heat was measured using a relaxation method on PPMS (Quantum Design) under 0 and 9 T.

## 3. Results and discussion

Herein, we report the discovery of a HEA-type superconductor $Co_{0.2}Ni_{0.1}Cu_{0.1}Rh_{0.3}Ir_{0.3}Zr_2$ with $T_c$ = 8 K, at which the onset of the superconducting transition was observed in the temperature dependences of magnetization, electrical resistivity, and specific heat. The sample obtained through arc melting was silver-colored. The calculated sample-weight loss during the pelletizing and arc-melting processes was less than 1%; the obtained sample weighed 1.1167 g. The measured composition of the sample was $Co_{0.187}Ni_{0.097}Cu_{0.083}Rh_{0.328}Ir_{0.305}Zr_2$, which was calculated by fixing the Zr amount to 2. Although a slight deviation of the measured composition from the nominal starting composition was observed, the measured composition was close to the nominal value. Therefore, we called the sample $Co_{0.2}Ni_{0.1}Cu_{0.1}Rh_{0.3}Ir_{0.3}Zr_2$ using the nominal value in this paper. Note that all the transition metal compositions are in the range of 5–35%, which meets one of the criteria of the HEA. The calculated $\Delta S_{mix}$ for the $Tr$ site was 1.47$R$. As shown in Fig. S1 (supplemental data), the homogeneous distribution of $Tr$ elements was confirmed through elemental mapping, suggesting no local phase separation in the order of μm.

Figure 1(a) shows the powder XRD pattern and the Rietveld refinement results. There is no clear indication of the presence of impure phases. Refinement with a tetragonal $CuAl_2$-type ($I4/mcm$) model resulted in a reliability factor of $R_{wp}$ = 8.7%. The refined lattice constants were $a$ = 6.5006(2) Å and $c$ = 5.5320(3) Å. The atomic coordinates were Zr($x, y, z$) = (0.1639(3), 0.6659(3), 0) and $T$($x, y, z$) = (0, 0, 0.25). Schematic images of the refined crystal structure are shown in Figs. 1(b,c).

Figure 2(a) displays the temperature dependence of the susceptibility ($4\pi\chi$) for $Co_{0.2}Ni_{0.1}Cu_{0.1}Rh_{0.3}Ir_{0.3}Zr_2$. Large diamagnetic signals, corresponding to the emergence of superconductivity, were observed below 7.8 K. Susceptibility below 300 K is shown in Fig. S2(a) in the Supplemental Data, which exhibits no clear magnetic transition below 300 K. Figure 2(b) displays the magnetic field dependence of magnetization (*M-H*) at 2.5 K. It is evident from the data that superconducting currents are generated. The magnetic $J_c$ estimated from *M-H* is displayed in Fig. S2 (Supplemental Data). When comparing the data with that for $CoZr_2$ ($T_c \sim 6$ K for our sample), we note that $J_c$ at low magnetic fields is comparable between them, but $J_c$ for $Co_{0.2}Ni_{0.1}Cu_{0.1}Rh_{0.3}Ir_{0.3}Zr_2$ at higher fields is higher than $CoZr_2$. To further analyze the effect of the introduction of a HEA site to $TrZr_2$ and related systems, experiments using single crystals are needed. The inset figure shows an *M-H* plot at low fields, which confirms that the lower critical field ($\mu_0H_{c1}$) is ~ 20 mT. From these magnetization data, we confirmed that $Co_{0.2}Ni_{0.1}Cu_{0.1}Rh_{0.3}Ir_{0.3}Zr_2$ is a typical type-II superconductor.

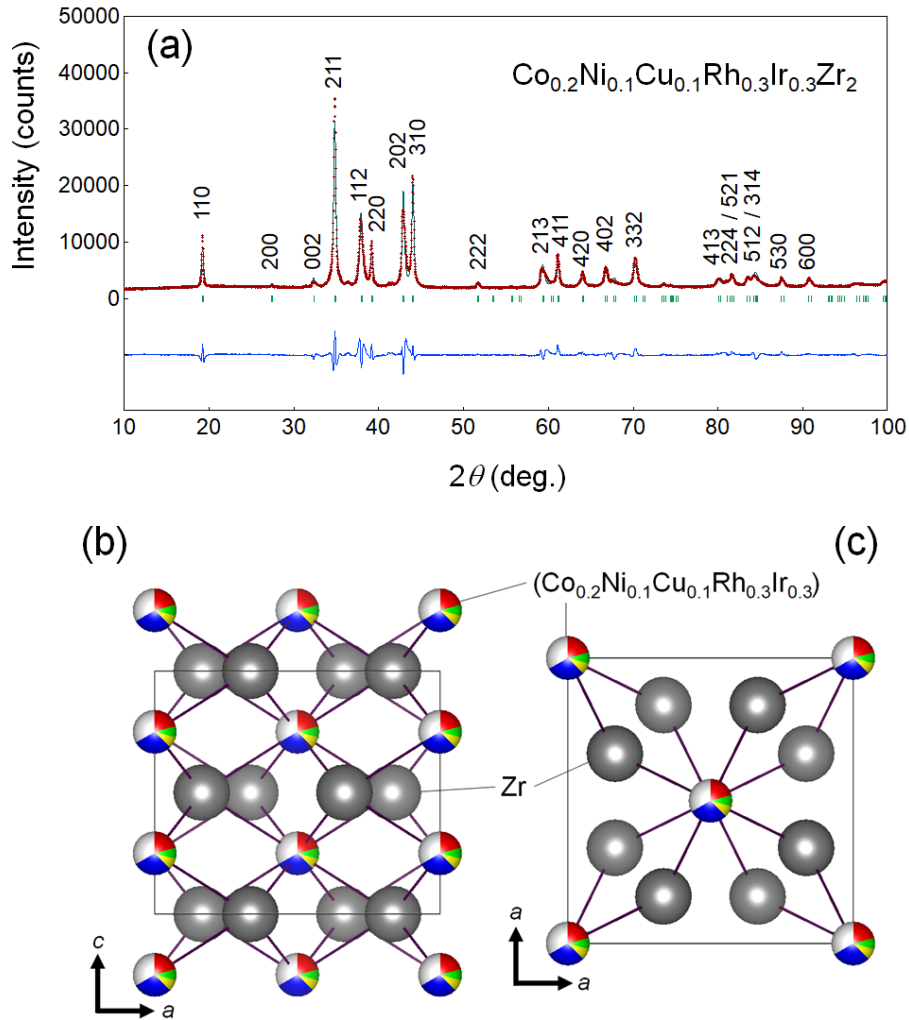

Fig. 1. (a) XRD pattern and Rietveld fitting for $Co_{0.2}Ni_{0.1}Cu_{0.1}Rh_{0.3}Ir_{0.3}Zr_2$. The numbers in the figure are Miller indices. (b, c) Schematic images of crystal structure of $CuAl_2$-type $Co_{0.2}Ni_{0.1}Cu_{0.1}Rh_{0.3}Ir_{0.3}Zr_2$.

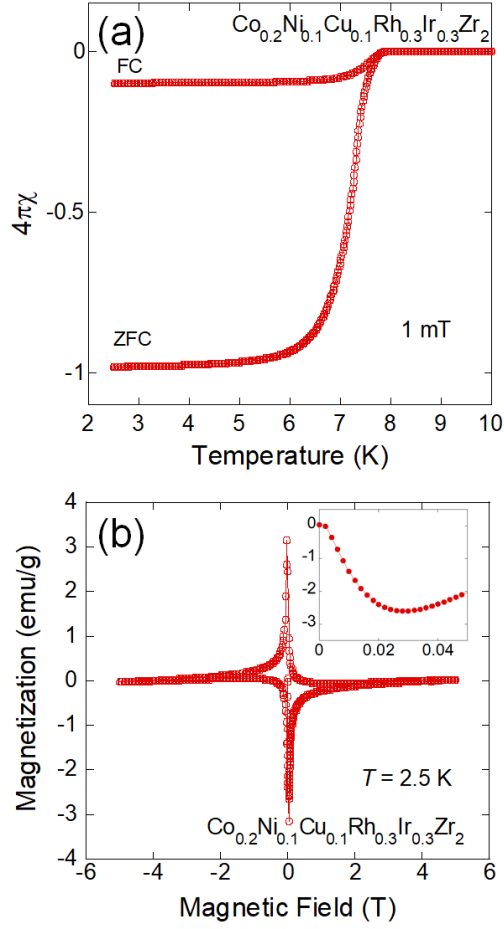

Fig. 2. (a) Temperature dependence of (ZFC and FC) susceptibility ($4\pi\chi$) for $Co_{0.2}Ni_{0.1}Cu_{0.1}Rh_{0.3}Ir_{0.3}Zr_2$. (b) Magnetic field dependence of magnetization for $Co_{0.2}Ni_{0.1}Cu_{0.1}Rh_{0.3}Ir_{0.3}Zr_2$ taken at $T = 2.5$ K. The inset shows low-field *M-H* data.

Figure 3(a) shows the temperature dependence of the electrical resistivity ($\rho$) for $Co_{0.2}Ni_{0.1}Cu_{0.1}Rh_{0.3}Ir_{0.3}Zr_2$. Typical metallic behavior was observed, which is consistent with previous reports on $TrZr_2$ compounds [27,33]. As shown in Fig. 3(b), the onset temperature ($T_c^{onset}$) and the zero-resistivity temperature ($T_c^{zero}$) were 8.0 and 7.5 K, respectively, at 0 T. $T_c$ decreased with increasing magnetic field [Fig. 3(b)]. To estimate the upper critical field at 0 K [$\mu_0H_{c2}(0)$], the resistive midpoint $T_c$ ($T_c^{\rho}$) and magnetic $T_c^M$ estimated from magnetization under various magnetic fields (see supplemental data, Fig. S3) are plotted in a magnetic field-temperature phase diagram in Fig. 3(c). Using the WHH model (Werthamer-Helfand-Hohenberg model) [34], which is applicable for a dirty-limit type-II superconductor, the $\mu_0H_{c2}(0)$ was estimated as 12 T.

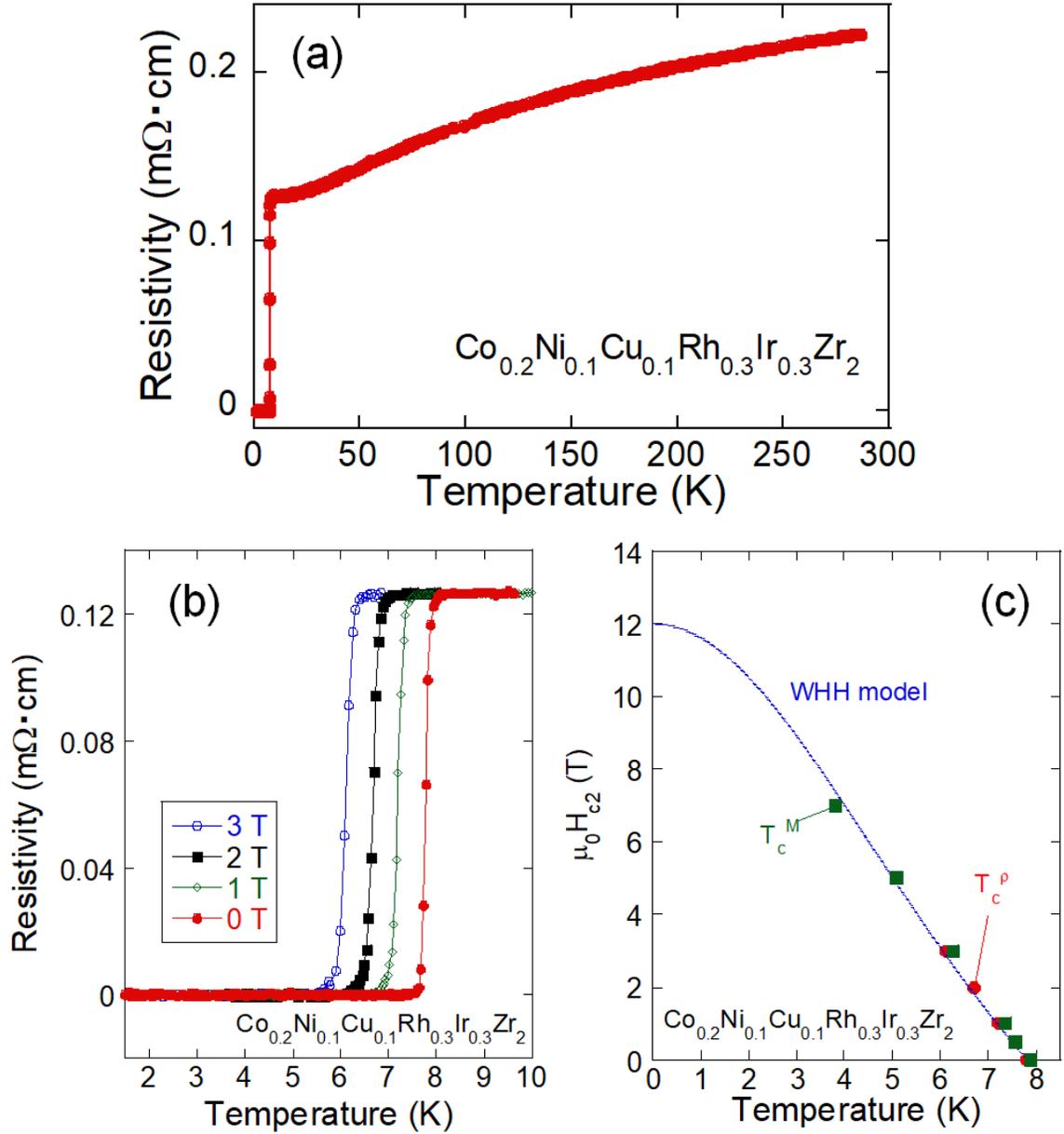

Fig. 3. (a) Temperature dependences of electrical resistivity for $Co_{0.2}Ni_{0.1}Cu_{0.1}Rh_{0.3}Ir_{0.3}Zr_2$. (b) Temperature dependences of resistivity for $Co_{0.2}Ni_{0.1}Cu_{0.1}Rh_{0.3}Ir_{0.3}Zr_2$ under magnetic fields. (c) Magnetic field-temperature phase diagram for $Co_{0.2}Ni_{0.1}Cu_{0.1}Rh_{0.3}Ir_{0.3}Zr_2$.

To confirm the bulk nature of the observed superconductivity, the specific heat was measured on a small piece (4.384 mg) of $Co_{0.2}Ni_{0.1}Cu_{0.1}Rh_{0.3}Ir_{0.3}Zr_2$. Figure 4(a) shows the temperature dependence of specific heat under 0 and 9 T. A clear jump was observed below 8.0 K under 0 T, but no superconducting transition was observed at 9 T. Therefore, we used the data under 9 T to estimate the electronic specific heat coefficient ($\gamma$) and the coefficient for the lattice specific heat contribution ($\beta$). The estimated $\gamma$ and $\beta$ were 20.7 mJ/mol·K$^2$ and 0.519 mJ/mol·K$^4$, respectively The Debye temperature ($\theta_D$) was estimated to be 224 K, which is close to that reported for a CoZr$_2$ single crystal [33]. To characterize the superconducting properties, the electronic contribution ($C_{el}$) at 0 T, which was calculated using $C_{el} = C - \beta T^3$, is plotted in the form of $C_{el}/T$ as a function of temperature in Fig. 4(b). The clear jump of $C_{el}/T$ and decrease in $C_{el}/T$ at low temperatures suggest bulk nature of the observed superconducting transition. However, the superconducting transition seen from $C_{el}$ was relatively broad compared to single crsytal data for CoZr$_2$ [33]. The origin of the broad transition is not clear, but the present data are sufficient to confirm the emergence of bulk superconductivity in the examined sample. The superconducting jump in $C_{el}$ ($\Delta C_{el}$) estimated at $T_c = 7.4$ K is $1.12\gamma T_c$ (see supplemental data Fig. S4), which is smaller than the value expected from the BCS model [35]. The smaller jump and the nature of the broad transition may imply the possibility of the semi-continuous evolution of the superconducting gap caused by HEA effects and/or the unconventional nature of the superconducing gap. To obtain further information about superconducting characteristics, specfic heat experiments at lower temperatures, scanning tunneling microscopy (STM), and angle-resolved photoemission spectroscopy (ARPES) are needed.

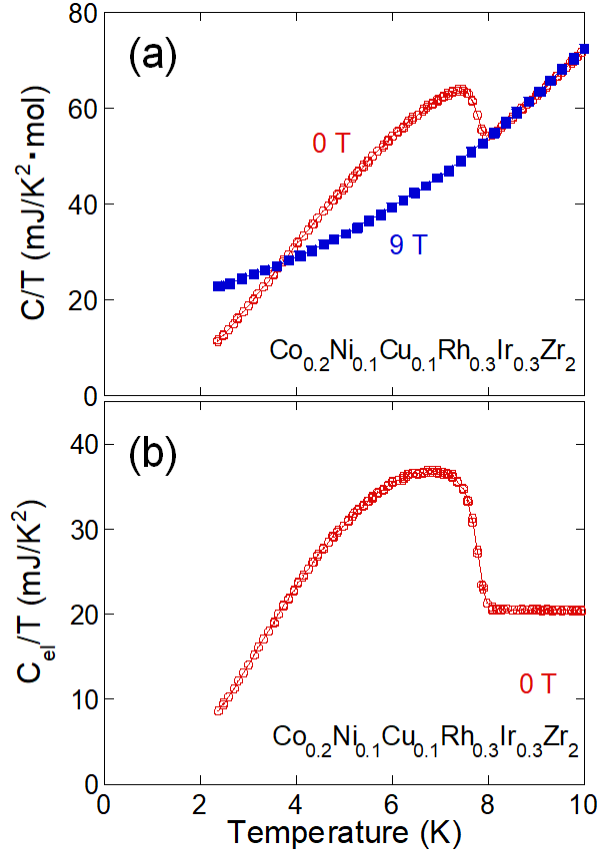

Fig. 4. (a) Temperature dependence of $C/T$ for Co$_{0.2}$Ni$_{0.1}$Cu$_{0.1}$Rh$_{0.3}$Ir$_{0.3}$Zr$_2$ at 0 and 9 T. (b) Temperature dependence of electronic specific heat $C_{el}/T$ at 0 T.

To discuss the effects of the HEA states at the $Tr$ site of $Tr\text{Zr}_2$ on the superconducting properties, we compared the $T_c$s of pure $Tr\text{Zr}_2$ and $\text{Co}_{0.2}\text{Ni}_{0.1}\text{Cu}_{0.1}\text{Rh}_{0.3}\text{Ir}_{0.3}\text{Zr}_2$. In known HEAs, properties or performances are sometimes close to the average of all the constituent metals. However, they sometimes result in unexpectedly high or low values [17], which is called the cocktail effect. In the case of $T_c$ of $\text{Co}_{0.2}\text{Ni}_{0.1}\text{Cu}_{0.1}\text{Rh}_{0.3}\text{Ir}_{0.3}\text{Zr}_2$, $T_c$ = 8.0 K appears to be close to the composition-weighted average value of $T_c$s for $Tr$ = Co, Ni, Rh, Ir. Therefore, a cocktail effect-like improvement of $T_c$ was not found in the present study. However, the averaging effect on $T_c$ is interesting from the electronic-state point of view. The electronic states of $\text{CoZr}_2$ were reported in Ref. 33. The density of states near the Fermi energy ($E_F$) is mainly composed of contributions of Co-d and Zr-d electrons. In addition, we compared the density of states near $E_F$ for $Tr$ = Co, Ni, Rh, Ir using the CompES-X, NIMS database [36]. Although the Zr-d contributions become higher for $Tr$ = Rh, Ir, the $Tr$-d and Zr-d contributions are essential for the density of states near $E_F$. Therefore, the disorder at the $Tr$ site does not directly suppress $T_c$ in this system, and we propose that the HEA-type $Tr\text{Zr}_2$ system is suitable for discussing the effects of HEA site in superconductors.

Here, we briefly describe the future prospects of the HEA-type $Tr\text{Zr}_2$ compounds. First, we expect that we can flexibly design various kinds of HEA-type $Tr\text{Zr}_2$ according to the elemental and compositional design described in the introduction part of this paper. In a superconductor database [26], $Tr\text{Zr}_2$-type superconductors containing $Tr$ = Sc, Fe, Co, Ni, Cu, Ga, Rh, Pd, Ta, Ir can be found. Therefore, the $Tr\text{Zr}_2$ phases with a HEA-type $Tr$ site will be useful for discussing the effects of long-range disorder, local phase separation, and/or modulation of local structure on superconducting properties. Furthermore, clarification of the HEA effects in superconductors will be important for the improvement of practical superconducting materials. Another possible merit of the HEA-type $Tr\text{Zr}_2$ is the congruent melting character of the $Tr\text{Zr}_2$ phases. As reported in Ref. 33, high-quality single crystals can be obtained through a simple crystal growth. Therefore, the HEA-type $Tr\text{Zr}_2$ phases will provide us with a platform to study the relationship between the superconducting properties and HEA states using single crystals.

## 4. Conclusion

We have reported the synthesis and superconducting properties of a new HEA-type superconductor $\text{Co}_{0.2}\text{Ni}_{0.1}\text{Cu}_{0.1}\text{Rh}_{0.3}\text{Ir}_{0.3}\text{Zr}_2$ with a HEA-type $Tr$ site. The composition of $\text{Co}_{0.2}\text{Ni}_{0.1}\text{Cu}_{0.1}\text{Rh}_{0.3}\text{Ir}_{0.3}\text{Zr}_2$ was designed based on information of the crystal structure ($\text{CuAl}_2$-type) and $T_c$ of the $Tr\text{Zr}_2$ compounds from the NIMS database (SuperCon). A polycrystalline sample was prepared using pure metals through arc melting. A superconducting transition was observed at 8.0 K in magnetization, electrical resistivity, and specific heat measurements. From the magnetization and resistivity data under magnetic fields, the upper critical field $\mu_0H_{c2}(0)$ was estimated as 12 T. Specific heat data suggests that the superconductivity observed in the sample is bulk in nature. The estimated electronic specific heat coefficient and the Debye temperature were close to those reported for a CoZr$_2$ single crystal. Although a cocktail effect, which is an unexpected (non-average) improvements of the performance, was not observed on the $T_c$ of Co$_{0.2}$Ni$_{0.1}$Cu$_{0.1}$Rh$_{0.3}$Ir$_{0.3}$Zr$_2$, the discovery of superconductivity in HEA-type $Tr$Zr$_2$ should open up new avenues to explore novel HEA-type supercondcutors and investigations on the relationship between the mixing entropy, local structure, and superconductivity in HEA-type compounds.

**Declaration of interest statement**

The authors decleare no competing interests.

**Declaration of interest statement**

Experimental data are available via reasonable requests to the corresponding author.

**Acknowledgements**

The authors thank R. Tsubota, N. Nakamura, A. Yamashita, and O. Miura for their assistance with the experiments. This work was partly supported by JSPS KAKENHI [Grant Number: 18KK0076] and the Advanced Research Program under the Human Resources Funds of Tokyo [Grant Number: H31-1].

**Supplemental Data**

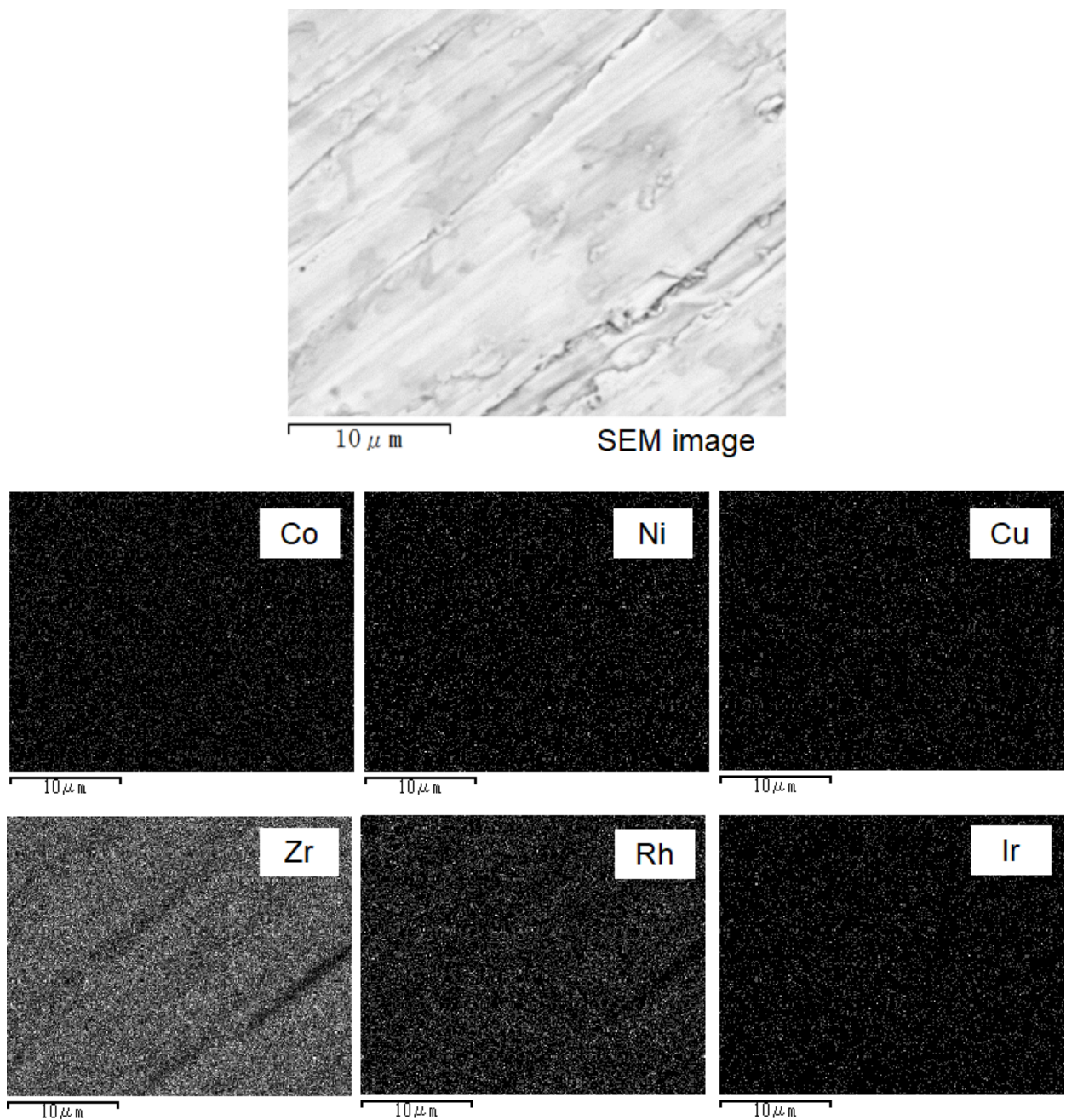

Fig. S1. SEM image and elemental mapping for $Co_{0.2}Ni_{0.1}Cu_{0.1}Rh_{0.3}Ir_{0.3}Zr_2$. The measurements were performed on TM3030 (Hitachi). The energy dispersive X-ray spectroscopy was performed by the Swift-ED system (Oxford).

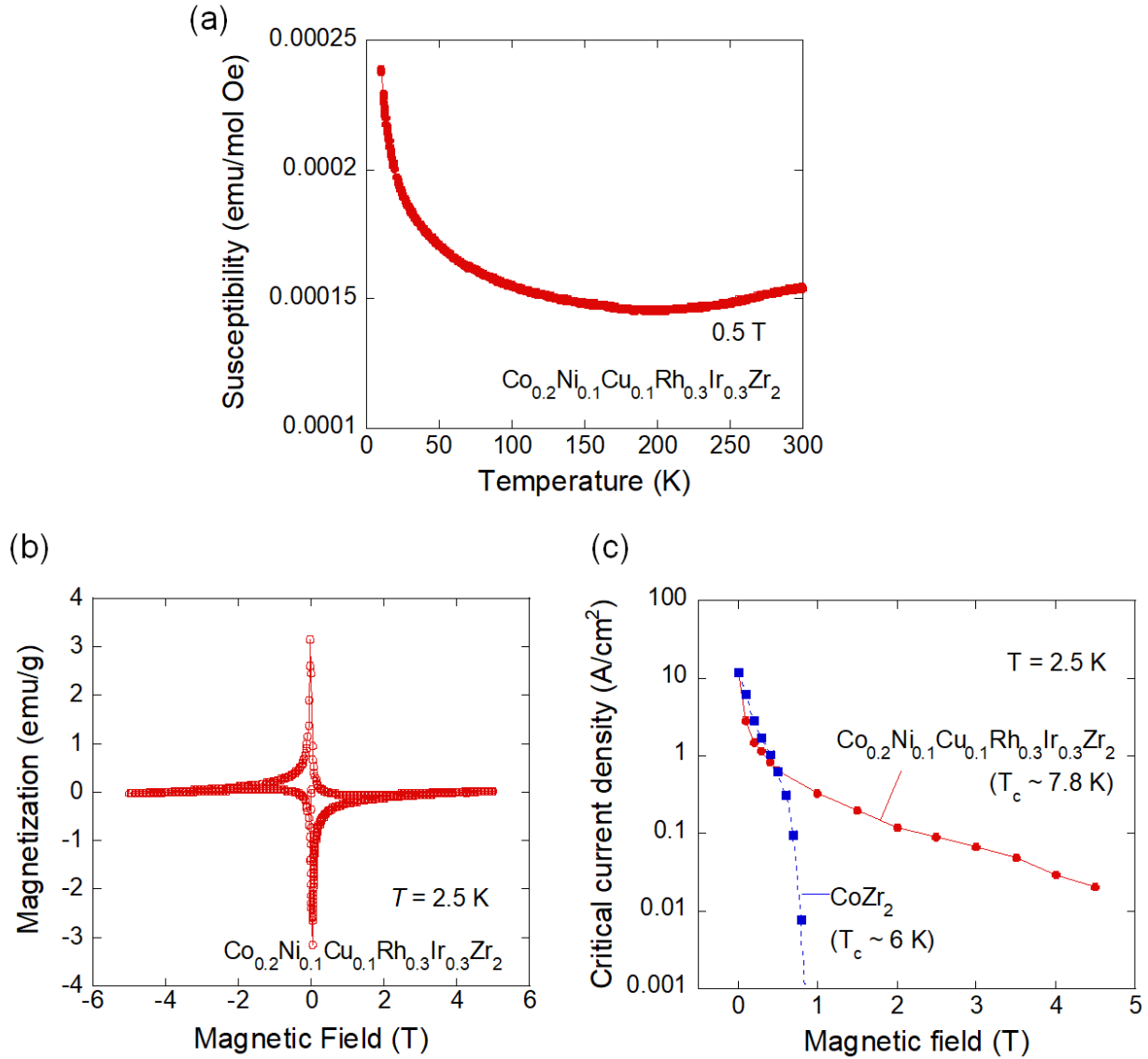

Fig. S2. (a) Temperature dependence of susceptibility for $Co_{0.2}Ni_{0.1}Cu_{0.1}Rh_{0.3}Ir_{0.3}Zr_2$. The molar susceptibility data was calculated using formula unit. (b) [same as Fig. 2(b)] Magnetic field dependence of magnetization ($M$-$H$ loop) for $Co_{0.2}Ni_{0.1}Cu_{0.1}Rh_{0.3}Ir_{0.3}Zr_2$. Although slightly-positive slope of magnetization was observed at higher fields, it is difficult to discuss about the origin with the data. Similar behavior was reported in $ZrZn_2$, an itinerant electron ferromagnet [S. Ogawa, N. Sakamoto, J. Phys. Soc. Jpn. 22, 1214 (1967)], but the molar susceptibility observed in the present sample is clearly lower than that for $ZrZn_2$. Furthermore, the HEA state in the $Tr$ site also makes the magnetic characteristics of the sample difficult. (c) Magnetic field dependence of magnetic $J_c$ (critical current density) estimated from Fig. S2(b). The magnetic $J_c$ was calculated by Bean's model [C. P. Bean, Rev. Mod. Phys. 36, 31 (1964)], which gives $J_c = 20\Delta M/b(1-b/3a)$ where $\Delta M$ is a width of the $M$-$H$ loop, and $a$ and $b$ are sample size. For comparison, data for $CoZr_2$ was plotted.

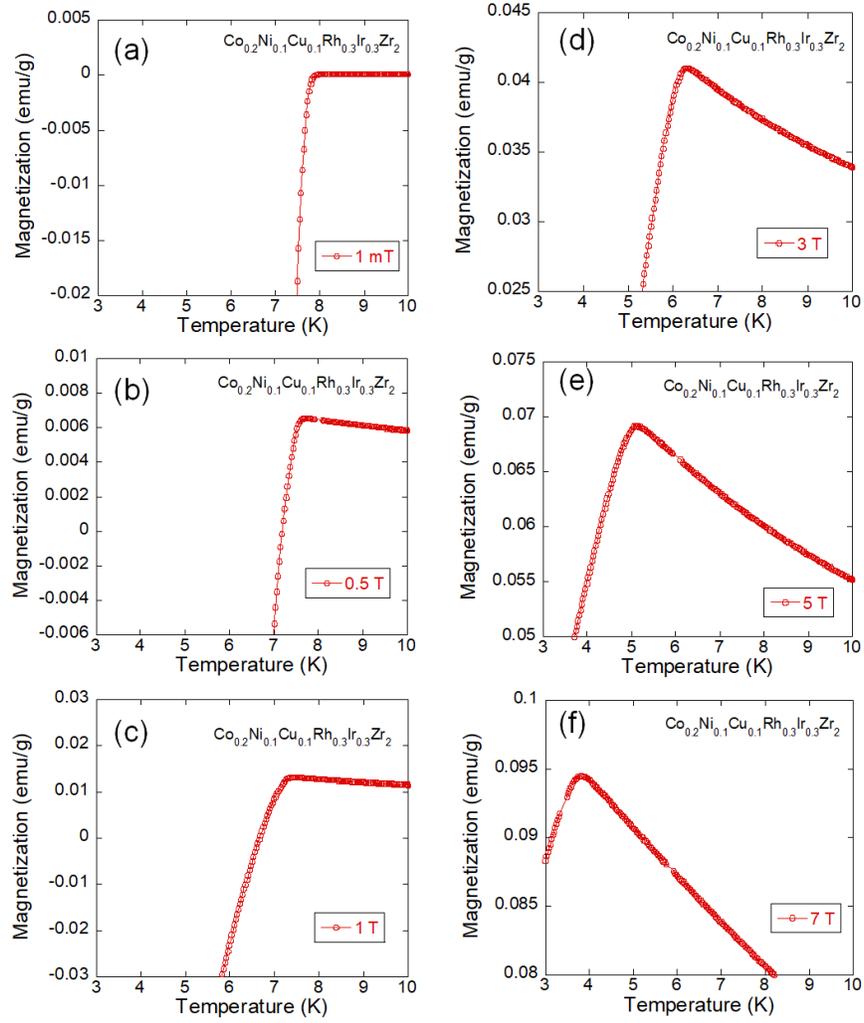

Fig. S3. Temperature dependences of magnetization for $Co_{0.2}Ni_{0.1}Cu_{0.1}Rh_{0.3}Ir_{0.3}Zr_2$ under various magnetic fields.

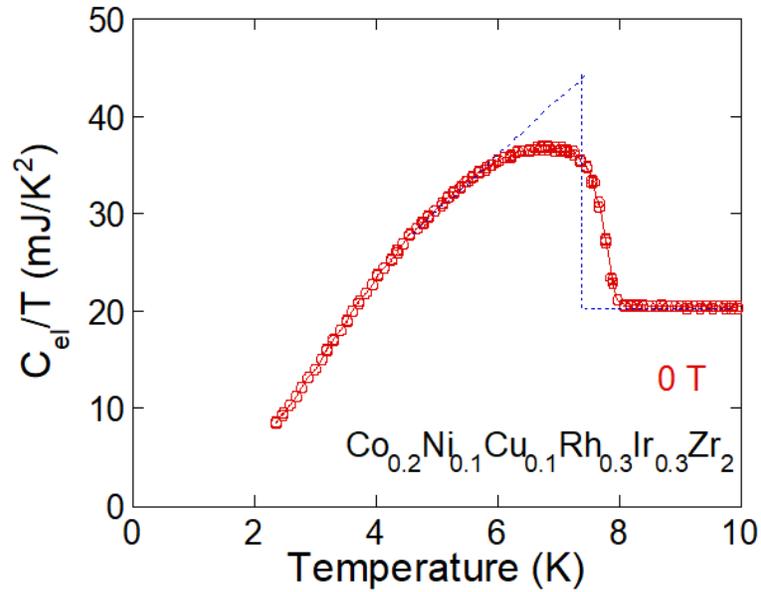

Fig. S4. Estimation of the electronic specific heat jump at $T_c$ (=7.4 K) for or $Co_{0.2}Ni_{0.1}Cu_{0.1}Rh_{0.3}Ir_{0.3}Zr_2$ under 0 T.

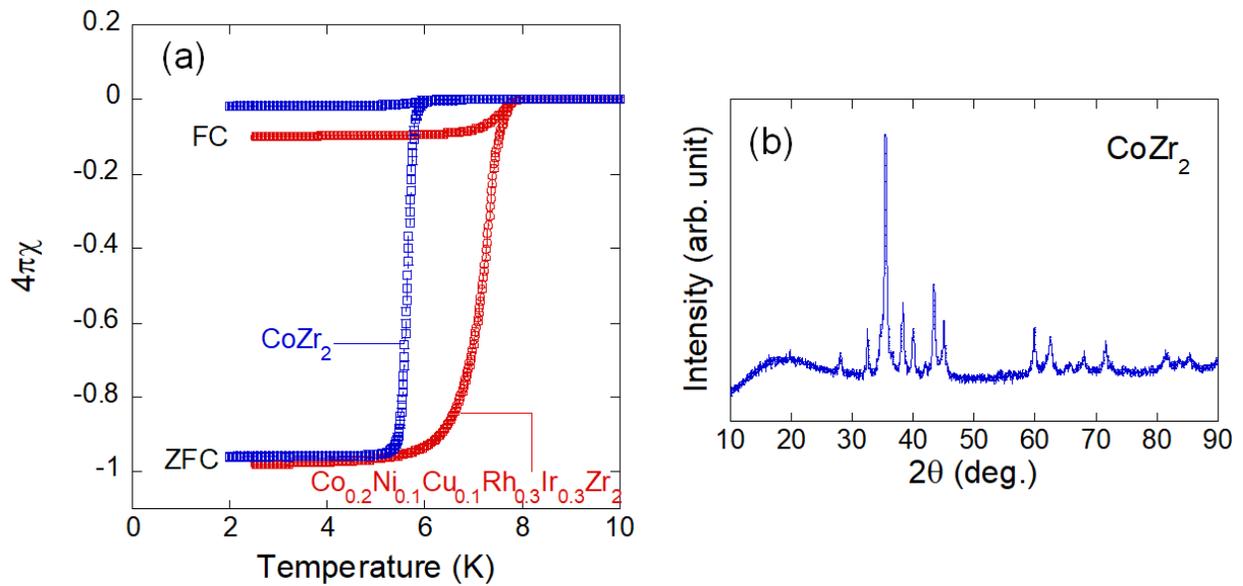

Fig. S5. (a) Temperature dependences of susceptibility for $CoZr_2$ and $Co_{0.2}Ni_{0.1}Cu_{0.1}Rh_{0.3}Ir_{0.3}Zr_2$. (b) XRD pattern for $CoZr_2$.